\documentclass[supercite,prb,amsmath,amssymb]{revtex4}
\usepackage{amsmath}
\usepackage{amssymb}
\usepackage{graphicx}% Include figure files
\usepackage{bm}% bold math
\def\degC{\,^\circ{\rm C}}

\def\eqnn#1{Eq.~(\ref{eq:#1})}

\def\figno#1{Fig.~\ref{fig:#1}}

\def\Im{{\rm \,Im}}

\def\sigk{\sigma_k}

\def\degC{$^\circ\,{\rm C}$}

\def\vev#1{\langle #1\rangle}
\def\Im{{\rm \,Im}}

\def\kb{k_{\scriptscriptstyle\rm B}}

\def\figW{80mm}
\begin{document}
\title{
  Dynamically enhanced optical coherence tomography
}
\author{Takahisa   Mitsui\footnote{E--mail:~{\tt      mitsui@phys-h.keio.ac.jp}.} and 
Kenichiro Aoki\footnote{E--mail:~{\tt ken@phys-h.keio.ac.jp}.} }
\affiliation{Research and Education Center for Natural Sciences and
  Dept. of Physics, Hiyoshi, Keio University, Yokohama 223--8521,
  Japan}
% \begin{abstract}
% \end{abstract}
% 1st para <150 w or so
\begin{abstract}
  In the investigations of inhomogeneous media, availability of
  methods to study the interior of the material without affecting it
  is valuable. Optical coherence tomography provides such a
  functionality, by providing depth resolved images of
  semi-transparent objects non-invasively. This is especially useful
  in medicine, and is used not only in research, but also in clinical
  practice. Optical coherence tomography characterizes each
  cross-section by its reflectance.  It is clearly desirable to obtain
  more detailed information regarding each cross-section, if
  available. We have developed a system which measures the fluctuation
  spectrum of all the cross-sections in optical coherence
  tomography. By providing more information for each cross-section,
  this can in principle be effective in tissue characterization and
  pathological diagnosis. The system uses the time dependence of the
  optical coherence tomography data, to obtain the fluctuation
  spectrum of each cross-section.  Additionally, noise reduction is
  applied to obtain the spectra without unwanted noise, such as
  shot-noise, which can swamp the signal. The measurement system is
  applied to samples with no external stimuli, and depth resolved
  thermal fluctuation spectra of the samples are obtained. These
  spectra are compared with their corresponding theoretical
  expectations, and are found to agree. The measurement system
  requires dualizing the detectors in the optical coherence
  tomography, but otherwise requires little additional equipment.  The
  measurements were performed in ten to a hundred seconds.
\end{abstract}
\maketitle 
\section{Introduction}
\label{sec:intro}
Optical coherence tomography (OCT)\cite{oct1,oct2} uses
  low-coherence reflectometry to obtain cross-sectional images of
  inhomogeneous media, such as biological tissue.  OCT is particularly
  useful in the biomedical area, since the imaging can be performed
  non-invasively, and in a relatively short time. As such, OCT is not
  only used in research, but has proven to be effective in clinical
  practice, with the fields of its application continuing to
  grow\cite{octReview1}.  In OCT, essentially, the reflectance of each
  cross-section along the depth direction is obtained with high
  spatial resolution (${\sim10\,\mu}$m), revealing the structure
  of the sample.  

In implementing OCT, two main approaches exist, referred to as
time-domain OCT (TD-OCT)\cite{oct1,oct2} and Fourier-domain OCT
(FD-OCT)\cite{sdoct1,sdoct2}. OCT makes use of interferometry, and
TD-OCT resolves the depth by effectively moving the reference (or the
sample) in the interferometry, while FD-OCT spectrographically
resolves the depth information. FD-OCT has an advantage that the OCT
image can be obtained with a single exposure, with the sensitivity a
few hundred times better than that of TD-OCT, but gives rise to
virtual images unless further processing is performed. The relative
merits of the two approaches were analyzed, and a method for removing
virtual images has been found
\cite{Mitsui99,Leitgeb2003,Boer2003}.  OCT essentially measures
the reflectance of each cross-section, and this can lead to information
regarding the matter buried beneath the surface, when the material is
semi-transparent. Results from OCT can also be combined with Doppler
velocimetry, and has been applied to
angiography\cite{octDoppler,octReview1,octDoppler2,octDopplerReview}.
It is clearly desirable to obtain more information, in addition to its
reflectance, regarding each cross-section in OCT.
% In FD-OCT, the
% information in the the time domain is unused so that it can be tasked
% to another purpose. 

In this work, we have developed a system that combines OCT with
thermal fluctuation measurements, which extracts physical properties,
such as the viscoelastic characteristics, of each cross-section in the
depth direction. We demonstrate the efficacy of this system using
various samples.  Furthermore, the noise is reduced to below
shot-noise levels using averaged correlations, making it possible to
extract spectral information otherwise unobtainable. The measurement
system can in principle lead to tissue characterization and
pathological diagnosis.
% The
%   measurement system is non-invasive, and requires only a short time
%   for measurements, with no moving parts. 
%
\section{The theory underlying dynamically enhanced optical coherence tomography}
\label{sec:theory}
\begin{figure}[htbp]
  \centering
  \includegraphics[width=160mm,clip=true]{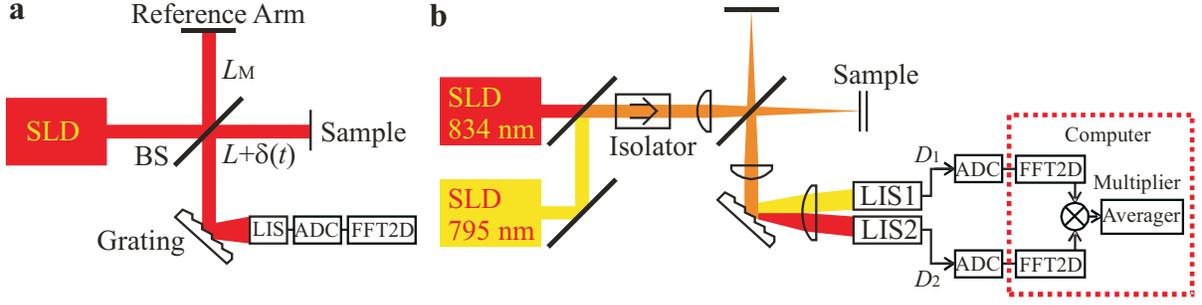}
  \caption{ {\bf Experimental setup for the dynamically enhanced OCT
      system:} {\bf a}, The basic conceptual setup of the measurement
    system:  Light from a superluminescent light diode (SLD), is fed
    into a Michelson interferometer with a reference arm, and a
    sample. The light from the interferometer is spectrally decomposed
    by a grating, and its intensity is measured for each
    wavelength by a linear imaging sensor (LIS). The photocurrent
    from LIS is digitized by an analog-to-digital converter (ADC), and
    Fourier transformed both in $k$, and $t$ (FFT2D), to obtain the
    power spectra of all the cross-sections in OCT. BS: beam splitter.
    {\bf b}, The schematics of the measurement system used in this
    work: The basic concept in {\bf a} is dualized with light sources
    with central wavelengths 795, 834\,nm, for noise reduction to
    observe at sub-shot-noise levels. Noise reduction is achieved by
    computing the averaged correlation of the two measurements
    (Averager). The light powers at the sample were 36\,$\mu$W,
    50\,$\mu$W, for the two wavelengths, respectively.}
  \label{fig:setup}
\end{figure}
In this work, we measure the time dependence of the FD-OCT signal to
obtain the thermal fluctuation spectra of each
cross-section in OCT. Measuring thermal fluctuation spectra of surfaces and
interfaces has proven to be effective in extracting the
physical properties of the material, such as  viscoelastic
characteristics\cite{ripplonExp,Cicuta2004,Sagis}.  
The basic concept underlying the measurement system is as follows:
Consider a Michelson interferometer with a reference arm, and a sample
with reflectance $R$, (\figno{setup}{\bf a}). For the light with a
particular wave number $k$, and angular frequency $\omega_{\rm L}$,
the detected light power, $P(k,t)$, from the interferometer is
\begin{equation}
  \label{eq:mic3}
  P(k,t)=A\left|e^{i2kL_{\rm M}-i\omega_{\rm L} t}+Re^{i2k(L+\delta(t))-i\omega_{\rm L} t}\right|^2=
  A\left[1+R^2+2R\cos2k(\Delta L+\delta(t))\right]
  \quad,
\end{equation}
where $L_M, L$ are the arm lengths of the interferometer to the
mirror, sample surface respectively, $\Delta L=L-L_{\rm M}$, and $A$
is a constant.  $\delta(t)$ is the fluctuation of the sample surface
location in the light direction, defined such that its average is
zero.
% only the time dependent part of this light power, which reduces to
% (\suppl)
% \begin{equation}
%   \label{eq:mic2}
%   P(k,t)\sim -4Rk\delta(t)\sin2 k\Delta L+\hbox{time independent terms}\quad,
% \end{equation}
% where $\Delta L=L-L_{\rm M}$.  
By Fourier transforming $P(k,t)$ {\it both} in $k$, and in $t$, the
power spectra of the fluctuations $S(\Delta L,f)$ for a given
$\Delta L$ is obtained, so that {\it all} the cross-sections are
computed simultaneously.
%  We note that virtual
% images do not appear, since only the time dependent parts of
% $P(k,t)$ are used.
%   Here, $k$ in the
% prefactor was assumed to be a constant, since 

In practice, the Fourier transform in $k$ is performed within a finite
region, determined by the spectral width of the incoming light.  This
gives rise to a form factor that selects the location of the
cross-section, $\Delta L$. %, and the height fluctuation spectrum.
Assuming a Gaussian spectrum for $k$, we obtain $Q(x,\omega)$, the
Fourier transform of $P(k,t)$ both in $k$ and $t$,
\begin{equation}
  \label{eq:sumK}
 Q(x,\omega)=\int_{-\infty}^{\infty}dt\, e^{i\omega t} \int _ {-\infty}^\infty {dk\over2\pi}\,
  {e^{-(k-k_0)^2/(2\sigk^2)}\over\sqrt{2\pi}\sigk}e^{ikx}P(k,t)\quad.
\end{equation}
Here, $\sigk$ is the standard deviation of
% Here, $\sigk={\rm FWHM}/(2\sqrt{2\ln 2})$ is the standard deviation of
the spectral distribution, and $\omega=2\pi f$.  Since we are
interested in time-dependent ($\omega\not=0$) fluctuations, time independent 
terms will be suppressed in the equations.  It should be noted that only
$\delta$ is time-dependent in $P(k,t)$.  The power spectrum of
the fluctuations at $x$ is $|Q(x,\omega)|^2$, up to a
constant.

Using $P(k,t)$ in \eqnn{mic3}, the integration over $k$ can be
performed to obtain
\begin{equation}
  \label{eq:qEq}
  Q(x,\omega)={AR\over2\pi}\int_{-\infty}^{\infty}dt\, e^{i\omega t} 
  \left[e^{-\sigk^2(x-2\Delta L-2\delta)^2/2+i k_0(x-2\Delta L-2\delta)}+
    e^{-\sigk^2(x+2\Delta L+2\delta)^2/2+i k_0(x+2\Delta L+2\delta)}\right]
\end{equation}
In this work, we shall concentrate on thermal fluctuations, which are at
the atomic scale, so that $k_0\delta\ll1$. Keeping the leading terms
in $\delta$, and we find
\begin{equation}
  \label{eq:delta}
  Q_+(x,\omega)=  {AR\over\pi}e^{-\sigk^2(x-2\Delta L)^2/2}
  \left[ik_0+\sigk^2(x-2\Delta L)\right]\tilde\delta(\omega)
  \quad,
\end{equation}
where tilde denotes the Fourier transform in $t$. We have selected the
term peaked in $x$ at $2\Delta L$ as $Q_+$. Another term exists with
$\Delta L\rightarrow -\Delta L$.  The power spectrum is obtained
essentially by squaring the above quantity,
% Consequently, the power spectrum is a
% Gaussian prefactor that selects out the spatial location times the
% power spectrum of height fluctuations there.
% This factor is similar to that of the standard
% OCT\cite{Swanson92,sdoct2,octReview1} , and leads to the depth
% resolution.
\begin{equation}
  \label{eq:qsq}
  \left|Q_+(x,\omega)\right|^2  = \left({AR\over\pi}\right)^2
  e^{-\sigk^2(x-2\Delta L)^2}
  \left[k_0^2+\sigk^4(x-2\Delta L)^2\right]\left|\tilde\delta(\omega)\right|^2
\end{equation}
So the doubly Fourier transformed power spectrum of $P(k,t)$ leads to a
thermal power fluctuation spectrum for each depth, $x/2$. 
Of the two terms in the square brackets in the above expression, the
first term is dominant unless the spectrum is broad. 
The Gaussian prefactor is similar to that in the standard
OCT\cite{Swanson92,sdoct1,octReview1} , except that it is squared in
the power spectrum used here.
%  The depth resolution is $1/(\sqrt2 \sigk)$,
% including this effect of squaring.
% , if the $\sin$ function is
% replaced by the $\cos$ function.

% where $L_M, L$ is the arm length of the interferometer to the mirror,
% sample surface respectively, and $\delta(t)$ is the fluctuation of the
% sample surface.  In this work, we shall only look at the time
% For fluctuations on a much smaller scale than the wavelength, as in
% this work, this can be reduced to
% \begin{eqnarray}
%   \label{eq:mic4}
%   P(k,t)&=&
%        A\left[1+R^2+2R\left(\cos2k\delta\cos 2k\Delta L 
%             - \sin 2k\delta \sin 2k\Delta L\right) \right]
%        \nonumber\\ &=& 
%                        A\left[1+R^2+2R\cos2 k\Delta L-2Rk\delta \sin (2k\Delta L)\right]\quad.
% \end{eqnarray}
% For the fluctuation spectra, only the time dependent part is used,
% which is essentially \eqnn{mic2}.
% The essence of the effects of the fluctuations in the light path to
% the sample surface under consideration is contained in the following
% simplified model.  

When the above procedure is applied to measuring the spectrum of
fluctuations inside a medium, two main additional points need to be
considered. First, $\Delta L$ is the optical length.  Second,
fluctuations from the medium in front of the measured location
contributes to the spectrum, in general, as we now explain. 
Let the cross-section under consideration be at a depth $d$ in an
uniform medium with an index of refraction, $n$.  Assume that the
surface of this medium is at distance $L'$ from the beam splitter, and
has fluctuation $\delta'(t)$ in the beam direction.  The light power
from the interferometer is,
\begin{equation}
  \label{eq:micSurf2}
  P'(k,t)=A\left|e^{i2kL_{\rm M}-i\omega t}
    +Re^{i2k(L'+\delta')+i2nk(d+\delta-\delta')-i\omega t}\right|^2\quad.
\end{equation}
Comparing this with the simple interferometer power, \eqnn{mic3}, one
can see that the effect is to replace $L$ by the optical length,
$L'+nd$, and to replace $\delta$ by
$n[\delta + (1-n)/n\times\delta']$.  The fluctuation $\delta'$ affects
the spectrum by changing the optical path length to the location of
interest. As such,  only  the fluctuations in the light path where the index
of refraction changes, or at partially reflecting surfaces, contribute to
the spectrum.  Also, it should be noted that the fluctuations $\delta$
is enhanced by $n$, while $\delta'$ is suppressed relatively by
$(1-n)/n$.  $\delta,\delta'$ are independent and the contribution to
the spectrum will be proportional to
$|\tilde\delta|^2+(n-1)^2/n^2{|\tilde\delta'|^2}$.
These fluctuations in the light path occur only at partially
reflective interfaces, and are included in the dynamically enhanced
OCT data for the smaller depth, so that they can be deconvoluted, in
theory.
% \paragraph{\bf Power spectrum and noise reduction}

In the current experiment, the height fluctuation power spectrum from
the Michelson interferometry, \eqnn{mic3}, is obtained. However, the
measured light powers $P_{1,2}$ at the two linear image sensors in the
setup \figno{setup}{\bf b} contain both the desired signal, $X$, along with
the unwanted noise $N_{1,2}$, as $P_j=X+N_j\ (j=1,2)$.  This noise
unavoidably contains the shot-noise, which is due to the quantum nature
of light, along with other experimental noise. We now briefly explain
how the noise reduction is applied is to obtain the spectrum.
The power spectrum of a signal, $X$, is obtained as
\begin{equation}
    \label{eq:spectrum}
{ S}(f) = \int_{-\infty}^\infty\!\! d\tau \,e^{-i2\pi f \tau} \vev{ X(t)
  X(t+\tau)} 
={1\over {\cal T}}\vev{\left|\tilde X(2\pi f)\right|^2}\quad,
\end{equation}
where $\vev{\cdots}$ denotes averaging, $f$ is the frequency, and
$\cal T$ is the measurement time.  In the experiment, the noise was
reduced statistically as follows\cite{MA1}: Let us assume that $N_j$
are {\it uncorrelated} between the two sensors.  Then,
\begin{equation}
    \label{eq:corr}
    \vev{\overline{\tilde P_1}\tilde P_2}\longrightarrow \vev{|\tilde
      X|^2}
    \qquad
    ({\cal N}\rightarrow\infty)\quad,     
\end{equation}
where $\cal N$ is the number of averagings.  Shot-noise is a typical
example of such uncorrelated noise, and it should be noted that this
averaging procedure also reduces any other extraneous noise that is
uncorrelated in the two photocurrents.  This reduction is statistical,
so the residual noise is $1/\sqrt{\cal N}$, relatively. 
% Since the
% time it takes to make a single measurement is $1/\Delta f$, where
% $\Delta f$ is the frequency resolution, the total measurement time is
% ${\cal N}/\Delta f$.
%
%  It should be noted that if the
% frequency resolution is kept constant relatively, the noise reduction
% factor is smaller at higher frequencies, which is convenient.

To understand the observed thermal fluctuation spectrum of each
cross-section in the dynamically enhanced OCT, the corresponding
theoretical spectrum needs to be understood.  The obtained
measurement, \eqnn{qsq}, is the height fluctuation power spectra for
each depth,  up to constants.
The height fluctuation spectrum measured using Michelson
interferometry is, theoretically\cite{MA2},
\begin{equation}
    \label{eq:spectrumMichelson}
    {S_{\rm h}}(f)=2\int_{k_{\rm min}}^\infty dk\,ke^{-w^{ 2}k^2/4}F(k,\omega)
\end{equation}
where $F(k,\omega)$ is the spectral function, $w$ is the beam waist,
and $k_{\rm min}$ is determined by the physical size of the
sample. For the results obtained below, we need the thermal
fluctuations of a liquid surface, whose spectral function
is\cite{Levich,Bouchiat,ripplonExp}
\begin{equation}
    \label{eq:ripplon}
    F(k,\omega)
    ={\kb T\over 4\pi}
    {\rho\over\eta^2k^3\omega}\Im\left[(1+s)^2+y-\sqrt{1+2s}\right]^{-1}
    % {k\tau_0^2\over \rho}\Im\left[(1+s)^2+y-\sqrt{1+2s}\right]^{-1}
    ,\quad
    s=  -i {\rho\omega\over 2\eta k^2}\quad
    % s\equiv  -i\omega\tau_0,\quad\tau_0\equiv {\rho\over 2\eta k^2}\quad
    y= {\sigma\rho\over4\eta^2 k},\quad.
\end{equation}
Here, $\rho,\sigma,\eta$ are the density, the surface tension and the
viscosity of the fluid. The above
theoretical considerations, physical properties of the fluid, and the
beam waist determine the theoretical fluctuation spectra completely.
% $k_{\rm min}=\pi/79\,\mu$m, $\pi/197\,\mu$m were used for water, oil,
% respectively.
\section{Experimental setup}
\label{sec:exp}
% The measurement system is shown in \figno{setup}, which combines the
% standard SD-OCT system with power spectra measurements for each
% surface, and a noise reduction system.
The basic concept of the measurement system, shown in
\figno{setup}{\bf a}, is as follows: The sample and the reference
constitute the arms of a Michelson interferometer, which uses the
light from a superluminescent light diode, a low-coherence light
source.  The spectrum of the interference signal light is obtained by
using a grating, and a linear imaging sensor (LIS).  This realizes the
FD-OCT measurement system. We further measure the time dependence of
the fluctuations of the light intensity, and use this to obtain the
power spectra of all the cross-sections in the depth direction,
individually.  The thermal fluctuations are at atomic scales, and
shot-noise, which inevitably arises due to the quantum nature of
light, can contribute significantly to their spectra.  To overcome
this problem, the setup (\figno{setup}{\bf b}) incorporates noise
reduction, whose theory was explained in the previous section. This is
achieved statistically by dualizing the light source and their
corresponding reflection measurements, then averaging the
correlation of the light powers measured at LIS's\cite{MA1}.
%  The total measurement time is
% for obtaining a spectrum is proportional to the number of averagings,
% and is under 100\,s in this work.

The technical details of the experimental setup are now explained: In
the setup, \figno{setup}, superluminescent diodes, QSDM-790-9, and
QSDM-830-9 (both QPhotonics, USA), with central wavelengths 795\,nm,
and 834\,nm and spectral widths 43\,nm, and 24\,nm, respectively,
were used as light sources.  This leads to a depth resolution of
$9\,\mu$m\cite{octReview1}.
%13 w/o 1/sqrt(2)
%
The objective lens used to focus the light onto the sample had a
numerical aperture of 0.015, and the beam waist at the sample was at
the diffraction limit, 20\,$\mu$m.  Light powers were 36, 50\,$\mu$W
at the sample, for the two aforementioned light sources,
respectively. For reflected light power measurement, linear image
sensors, S12198-512Q (Hamamatsu, Japan) were used.  14 bit
analog-to-digital converters, ADXII-14 (Saya, Japan) were used to
digitize the photocurrent.  Fourier transforms and averagings were
performed on a computer.
The clock for the linear image sensor is 10\,MHz, and 1024 cycles are
necessary for one readout, and 1024 lines were measured in a time series.
Therefore, one measurement requires $0.1\,$s. Typically a few
hundred to a thousand measurements were taken for the averaging, so
the time required is $100\,$s or less.

\section{Observations of depth resolved thermal fluctuation spectra}
\label{results}
\begin{figure}[htbp]
  \centering
\includegraphics[width=35mm,clip=true]{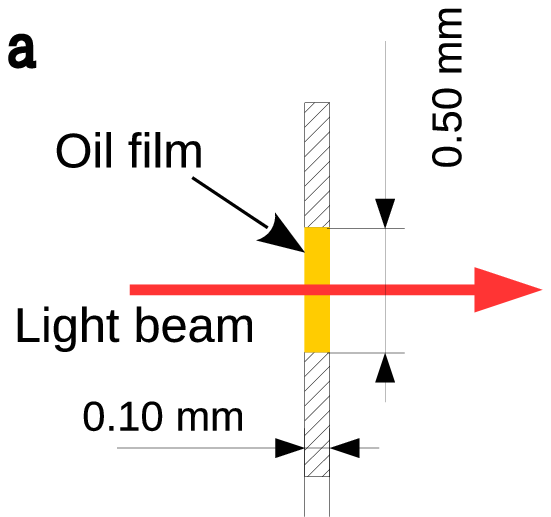}\hskip5mm
% 1.05 ave
\includegraphics[width=\figW,clip=true]{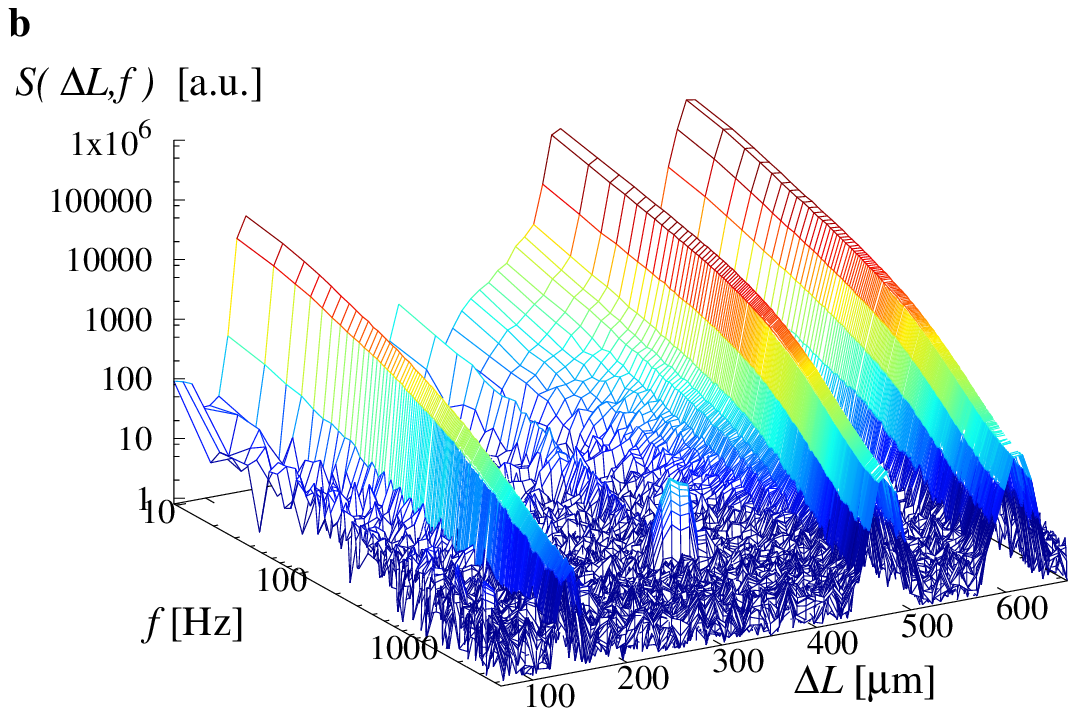}\\
\includegraphics[width=\figW,clip=true]{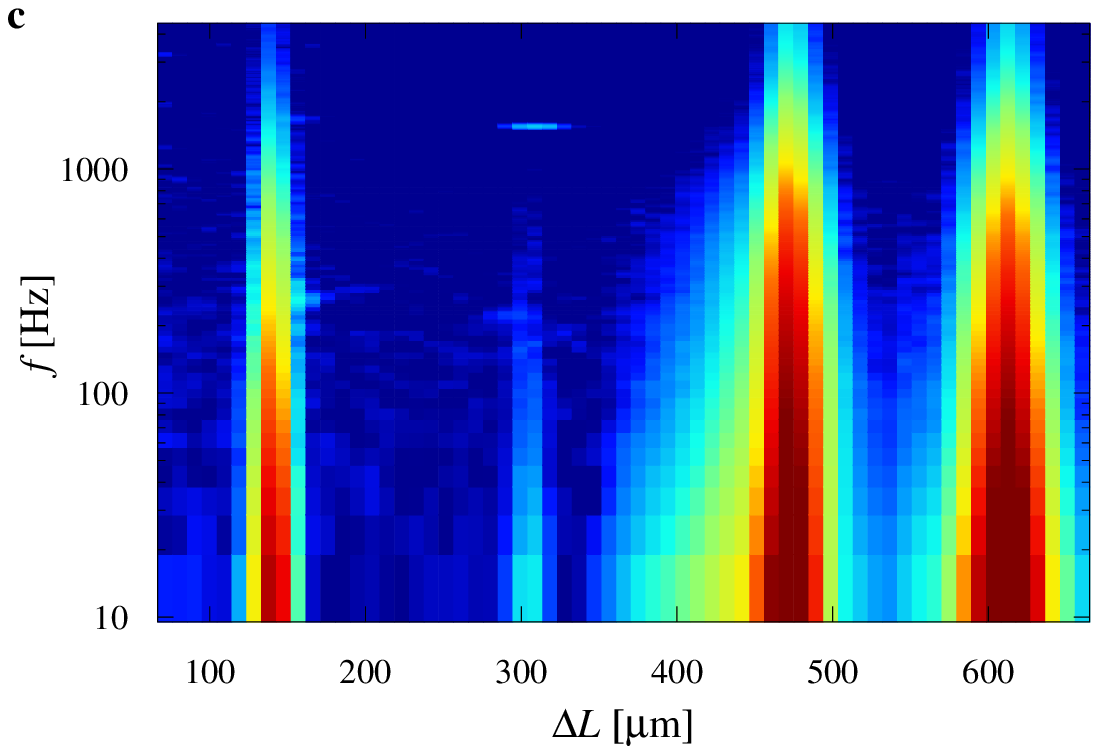}
\includegraphics[width=\figW,clip=true]{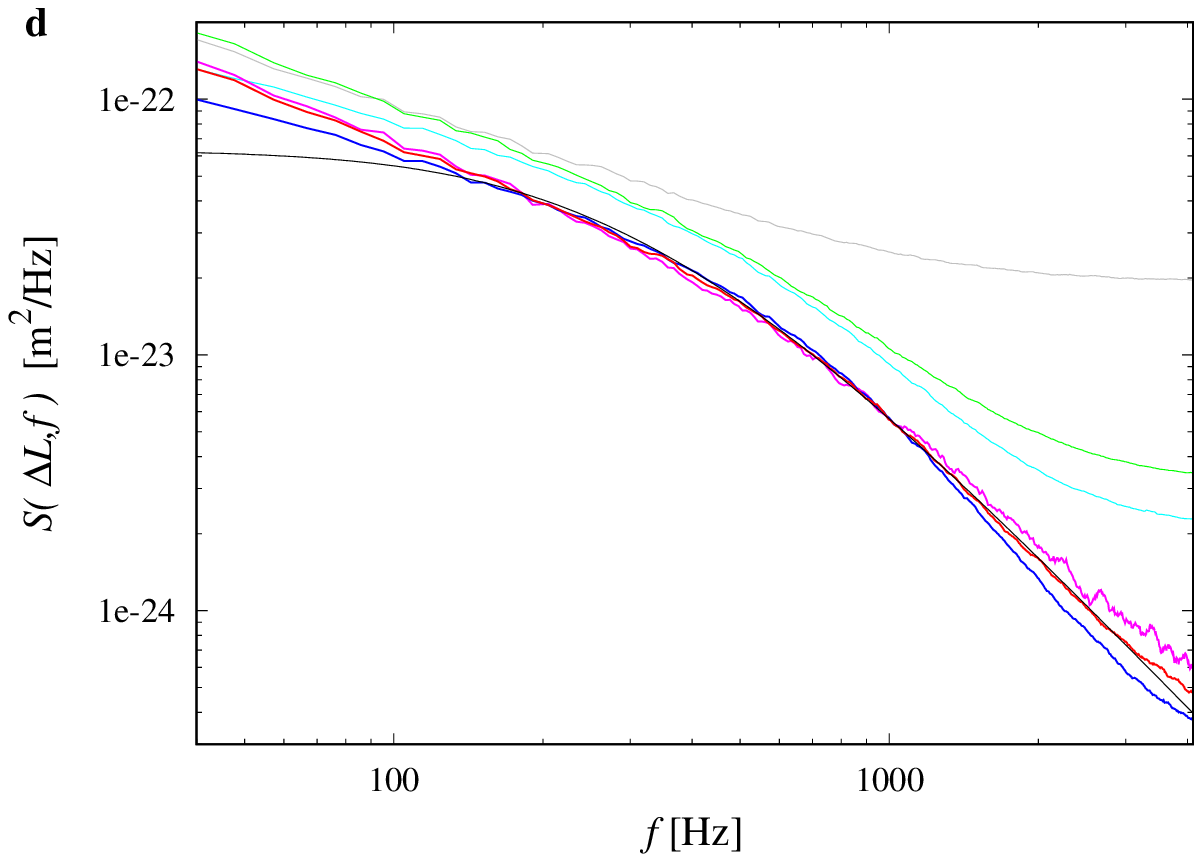}
\caption{{\bf Dynamically enhanced OCT measurements of an oil film:}
  {\bf a}, Oil is suspended perpendicularly to the light direction in
  a circular hole with diameter $0.50\,$mm, thickness $0.10\,$mm.
  {\bf b,c}, The thermal fluctuations of each cross-section along the
  light beam and the contour plot.  $f$: frequency. {\bf d}, The
  thermal fluctuation spectra of the two strongly reflective
  cross-sections at $\Delta L=470\,\mu$m (red), 613\,$\mu$m (blue),
  and the third reflective cross-section at $\Delta L=143\,\mu$m
  (magenta). These measured spectra all have essentially the same
  shape, and their theoretical prediction is shown as the black line,
  which agrees with the measurements quite well. The spectra without
  the statistical noise reduction are also shown for the three
  cross-sections, which includes the shot-noise (cyan, green, gray
  lines, respectively). The noise is relatively larger for smaller
  reflected power, and the observed shot-noise levels are consistent
  with theory.}
  \label{fig:oil}
\end{figure}
In \figno{oil}{\bf b,c}, the thermal fluctuation spectra for the
cross-sections of an oil film suspended in a hole are shown.  The
configuration of the experiment is shown in \figno{oil}{\bf a}, and
oil refers to silicone oil, Shin-Etsu KF96 300cs\cite{Shinetsu}, with
index of refraction 1.40, here, and below. Two strongly reflecting cross-sections
centered at $\Delta L=475,\ 618\,\mu$m can be seen, with another
signal peak at $\Delta L=143\,\mu$m.  The first two cross-sections
correspond to the two oil interfaces with air. The thickness 0.10\,mm
agrees with its optical depth, 0.14\,mm, the difference between the
two values of $\Delta L$.  The signal peak at $\Delta L=143\,\mu$m is
caused by the interference between the two cross-sections, and
$\Delta L$ is the optical depth of the film.
The thermal fluctuation spectra of these three cross-sections should
all correspond to that of the height fluctuations of the oil surface,
and they are essentially identical, as seen in \figno{oil}{\bf d}.
Using known physical properties of the oil, the corresponding
theoretical spectrum can be computed which is seen to agree quite well
with the measured spectra.  The physical properties of oil
$(\rho\,{\rm [kg/m^3]}, \sigma\,{\rm [kg/s^{2}]},\eta\,{\rm [kg/(
  m\cdot s)]})=(970,0.0211,0.291), (997,0.072,8.99\times10^{-4})$,
with the experimental temperature of 25\degC, were used here, and
below.  In the spectrum, $k_{\rm min}=\pi/(100\,\mu$m) was used,
considering the thickness of the oil film. There is a normalization
factor for each measured spectrum, which was chosen to match the
overall magnitude of the theoretical spectrum.  In \figno{oil}{\bf d},
the fluctuation spectra with noise reduction are
$ \vev{\overline{\tilde P_1}\tilde P_2}$, and the spectra without it
are $ \vev{\overline{|\tilde P_j}|^2}\ (j=1,2)$, in the notations of
the previous section, with the same normalizations.
The statistical noise reduction can be seen to be important for
obtaining the precise spectrum. While the shot-noise can also be
reduced relatively by increasing the light power, this will lead to
invasive measurements in general.
%  In
% the measurements in this work, the total light power shone on a sample
% was 86\,$\mu$W.

\begin{figure}[htbp]
  \centering
\includegraphics[width=40mm,clip=true]{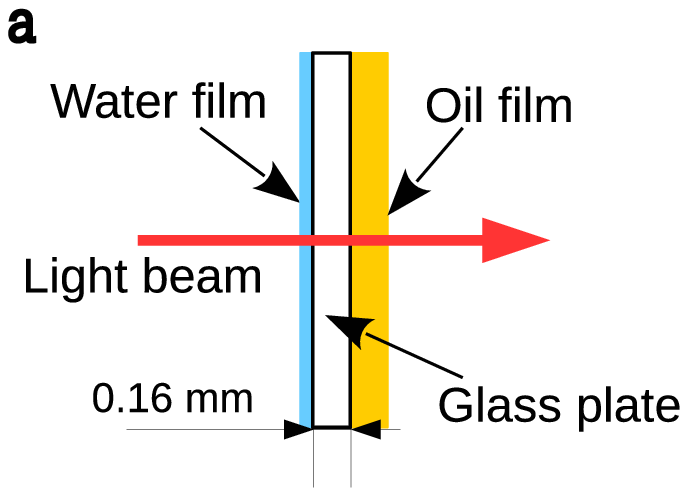}\hskip5mm
\includegraphics[width=\figW,clip=true]{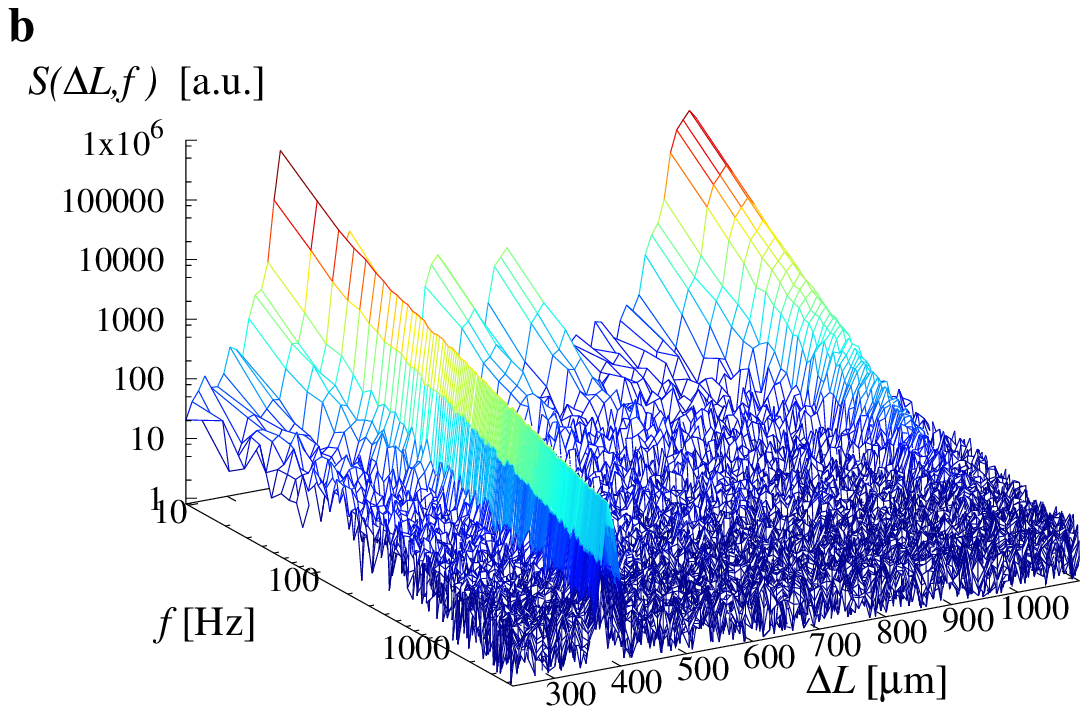}\\
\includegraphics[width=\figW,clip=true]{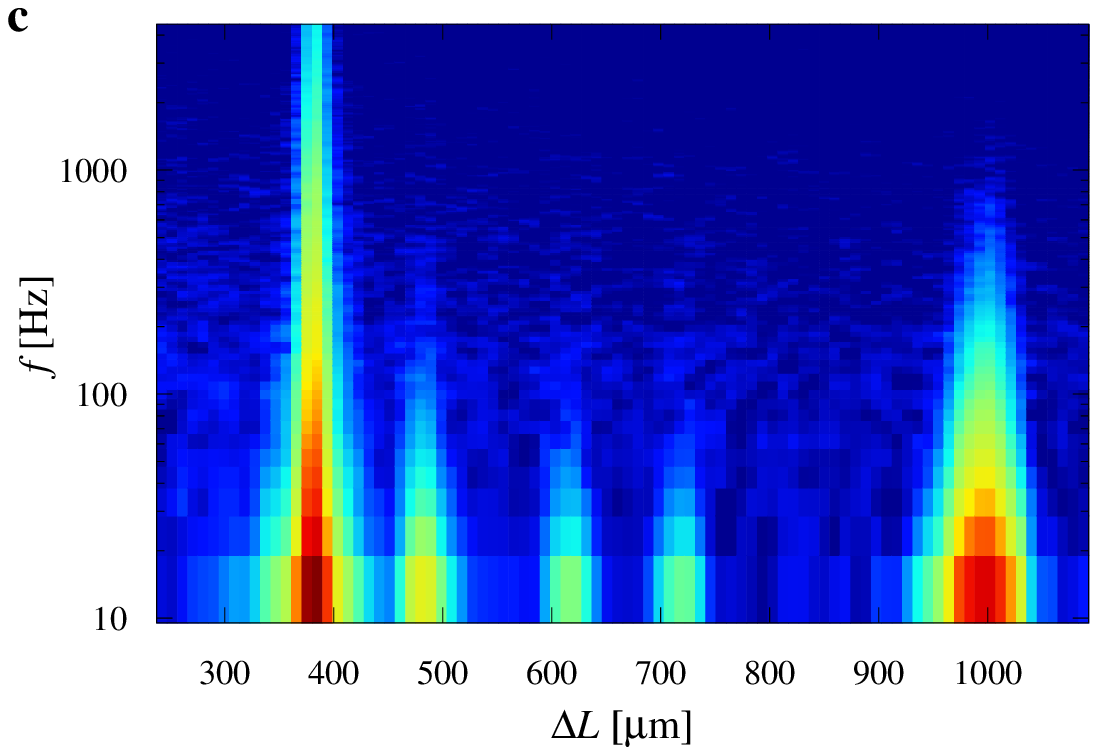}
% with Jackles formula
\includegraphics[width=\figW,clip=true]{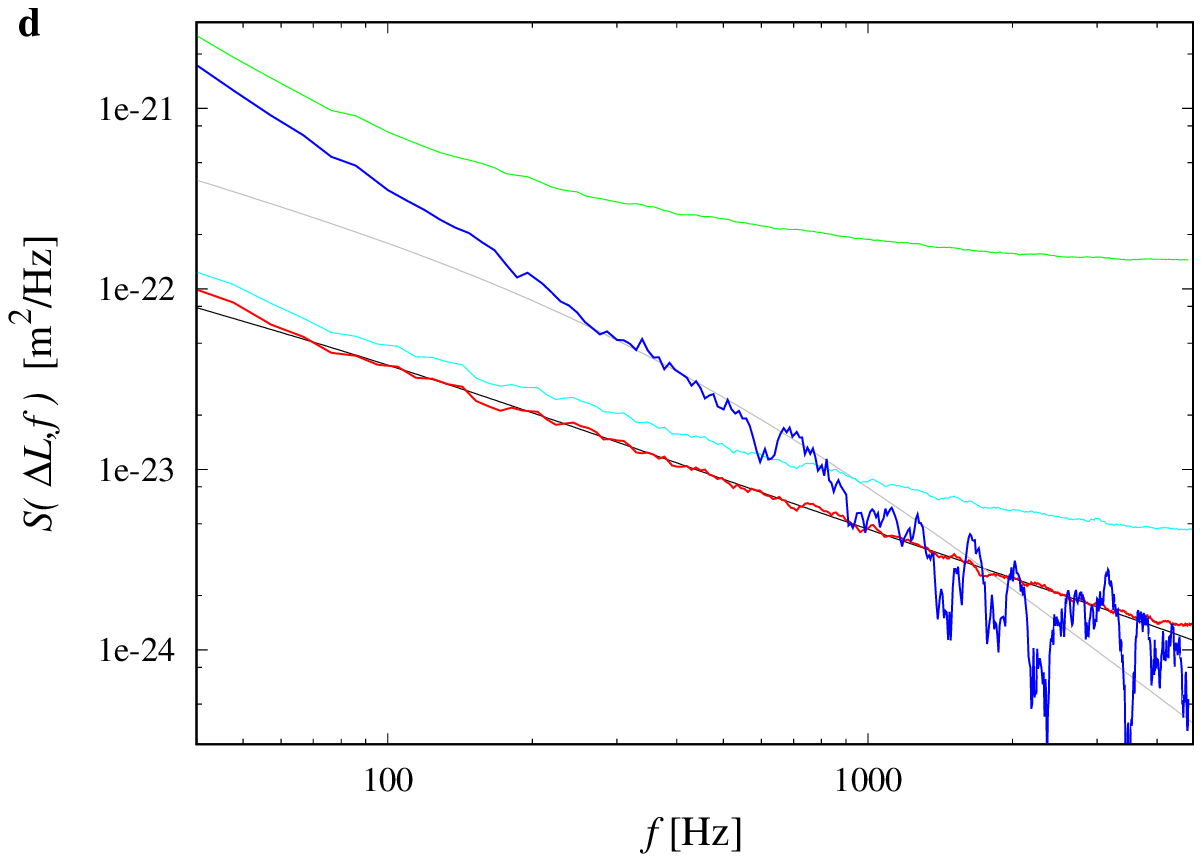}  
\caption{{\bf Dynamically enhanced OCT measurements of water and oil
    film fluctuations:} {\bf a}, Configuration: water, and oil films
  are perpendicular to the light direction, and are separated by a
  glass plate (thickness 0.16\,mm). {\bf b,c}, Thermal fluctuations of
  each cross-section in OCT, and the contour plot.  {\bf d}, The
  thermal fluctuation spectra for the two most reflective
  cross-sections in {\bf b,c}, at $\Delta L=380\,\mu$m (red), and
  998$\,\mu$m (blue). The theoretical spectra for water (black), oil
  (gray) are also shown and are consistent with the measurements. The
  spectra without the statistical noise reduction are also shown for
  the two cross-sections (cyan, green, respectively). }
  \label{fig:waterOil}
\end{figure}
In \figno{waterOil}{\bf b,c}, the measured thermal spectra of the
cross-sections of oil and water films, separated by a glass plate are
shown (configuration in \figno{waterOil}{\bf a}).  The structure of the
fluctuation spectra of the cross-sections, whose reflective peaks are
shown in \figno{waterOil}{\bf b,c} is substantially more complicated
than the previous case. The largest two peaks in \figno{waterOil}{\bf
  b,c} centered at $\Delta L=380, 998\,\mu$m can be identified as the
interfaces of water, and oil with air, respectively. The thermal
fluctuation spectra of these interfaces are shown in
\figno{waterOil}{\bf d}, with their corresponding theoretical spectra
with the corresponding boundary conditions\cite{Jackle} were used.
The physical properties of water used were\cite{CRC}
$(\rho\,{\rm [kg/m^3]}, \sigma\,{\rm [kg/s^{2}]},\eta\,{\rm [kg/(
  m\cdot s)]})=(997,0.072,8.99\times10^{-4})$.
The two spectra are seen to be quite distinct, and the thermal
fluctuation spectrum of the water surface agrees excellently with its
theoretical spectrum. For the oil surface, the measured and its
theoretical spectra are consistent, but there is a mismatch at lower
frequencies, which could be due to the effect of fluctuations in front
of it.  The spectra without the statistical noise reduction are also
shown, and it can be seen that the noise reduction is crucial for
extracting the spectral properties.  As in \figno{oil}, the overall
magnitudes of the spectra have been fitted to the theoretical spectra.
% The water interface fluctuation spectrum,
% which is in front, should contribute to the back side fluctuation
% spectra, but since the magnitude of the spectrum of water interface
% fluctuations are relatively small except at high frequencies, and is
% further suppressed (\suppl), the effect is not strongly visible in the
% oil interface fluctuations.  
The peaks centered at $485,\ 722\,\mu$m correspond to the front and
back surfaces of the glass plate. While the fluctuations of the glass
surfaces themselves are too small to be observable in this experiment, the water
surface fluctuations in front of them in the light path contribute to
the spectra, as explained in the previous section. The peak centered
at $\Delta L=618\,\mu$m in \figno{waterOil}{\bf c} corresponds to the
interference between the water and the oil surface, and $\Delta L$
corresponds to its optical length.  The physical thicknesses of the
water, oil films can be deduced to be $79, 197\,\mu$m respectively,
and the glass plate has a thickness of 0.16\,mm. These values explain
the values  $\Delta L$ of the peaks consistently.

\begin{figure}[htbp]
  \centering
\includegraphics[width=\figW,clip=true]{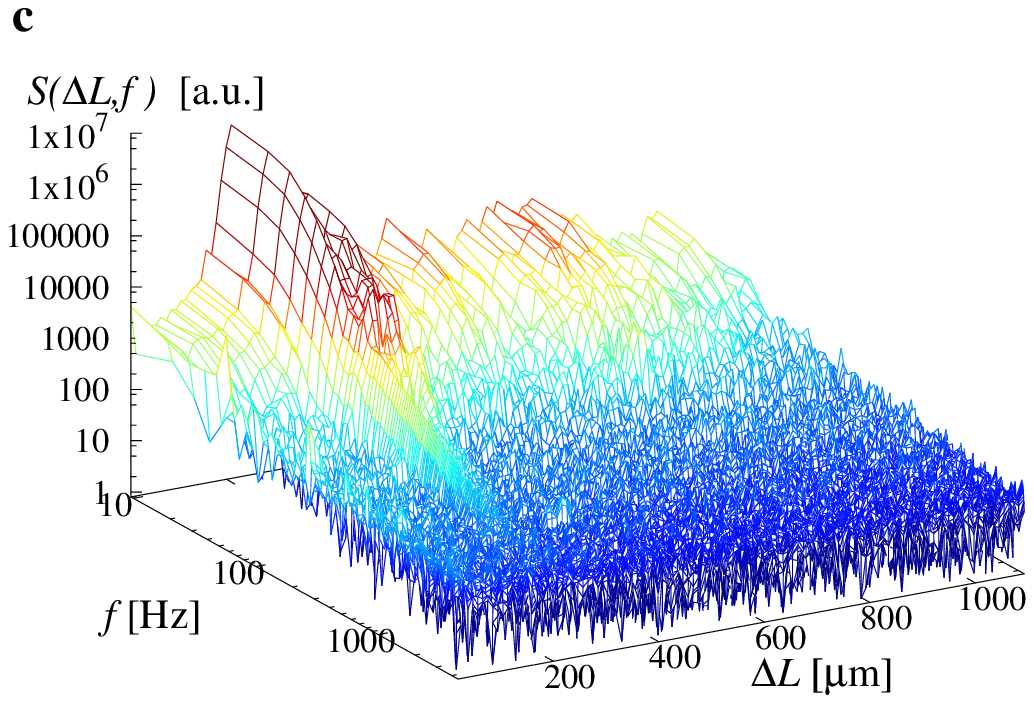}
\includegraphics[width=\figW,clip=true]{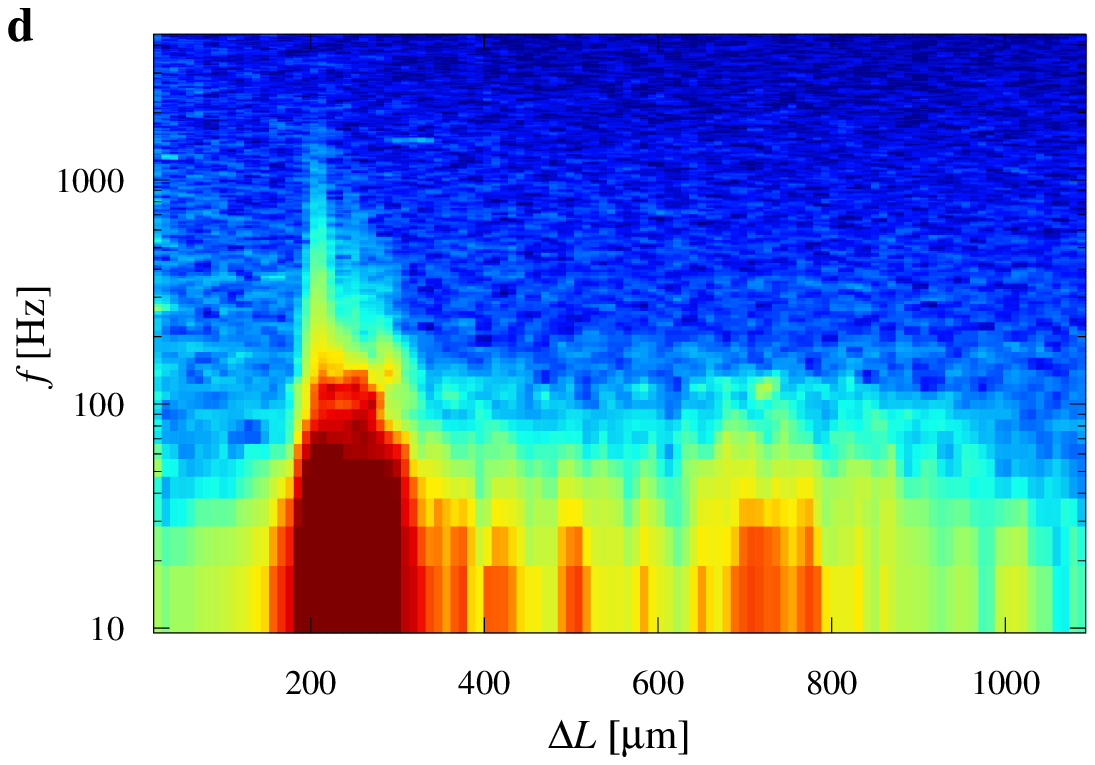}
\includegraphics[width=\figW,clip=true]{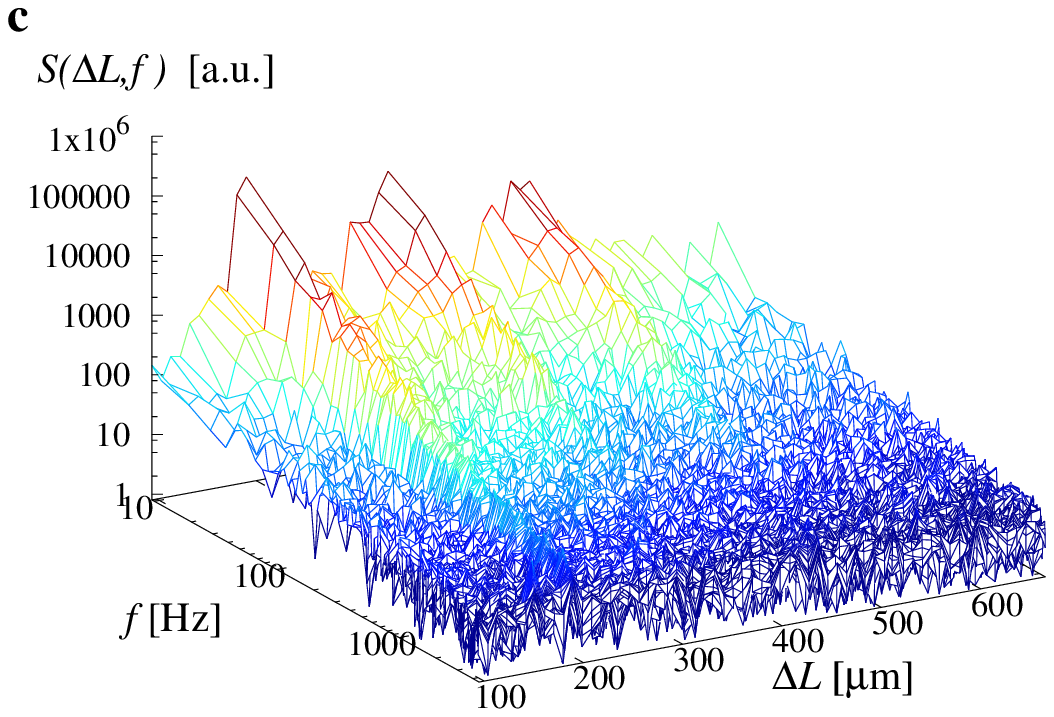}
\includegraphics[width=\figW,clip=true]{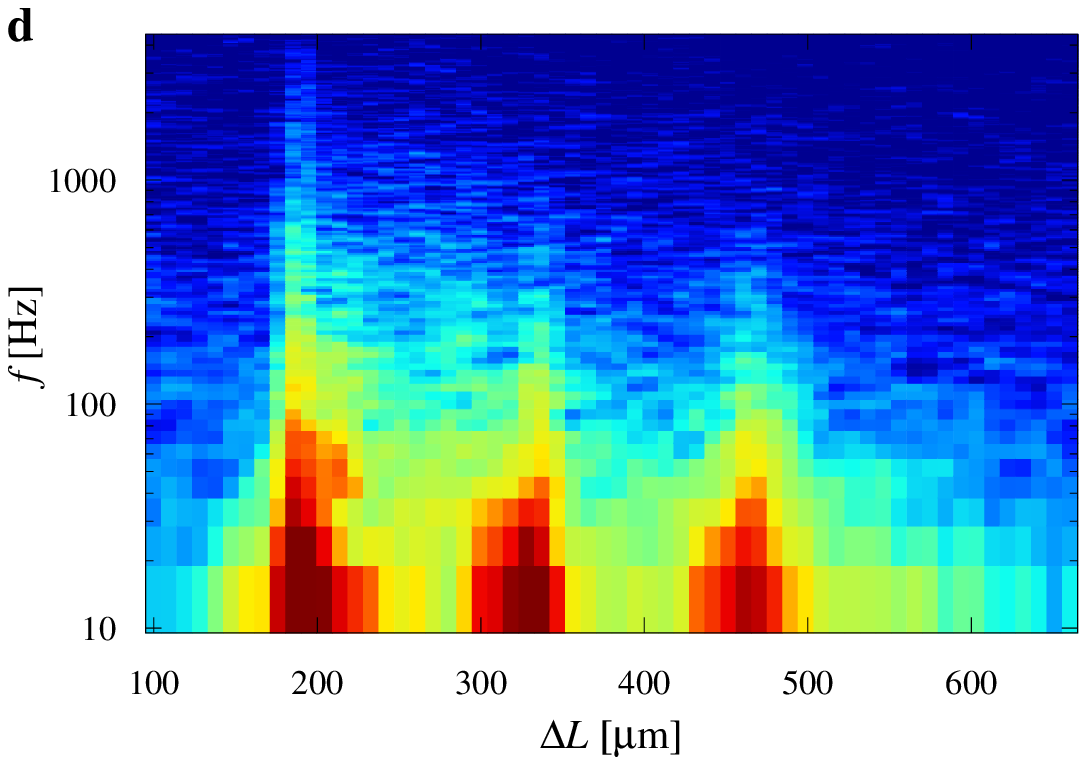}
\caption{{\bf Dynamically enhanced OCT measurements of thermal
    fluctuations of the structure of a finger and those of a sweetfish
    eye:} {\bf a,b}, Thermal fluctuation spectra for the cross-sections
  of a finger, and the contour plot. {\bf c,d}, Thermal fluctuation spectra of
  the cross-sections of a sweetfish eye, and the contour plot. }
  \label{fig:finger}
\end{figure}
In \figno{finger}{\bf a,b}, dynamical enhanced OCT images of a finger
(palm forward direction), and in \figno{finger}{\bf c,d} those of a
sweetfish eye are shown. Here, the media are quite inhomogeneous, and
the thermal spectra of reflective surfaces can be seen
continually. The observed spectra for the cross-sections do not seem
to correspond to the spectra of simple fluids like oil, water or
elastic materials, and more investigation is necessary to determine
their properties. The dependence of the sweetfish eye surface
fluctuation spectrum on the moisture level of the surface has been
observed previously\cite{MA1}, and the dependence of the depth
resolved spectra on moisture levels would be of interest.

\section{Summary and discussions}
\label{sec:disc}
In this work, we have developed a system dynamically enhancing OCT,
and demonstrated that thermal height fluctuations, hence the physical
properties of each cross-section in OCT, can be obtained this
way. Required extra instrumentation over OCT is minimal, while
essentially doubling the measurement apparatus if noise reduction is
incorporated. We restricted our attention to thermal fluctuations in
this work, which are spontaneous, and are at the atomic
scale. However, the system can be applied to motions, and also to
fluctuations that have been excited externally. Their power spectra
can have larger magnitudes, and consequently be easier to measure.
\section*{Acknowledgments }
K.A. was supported in part
by the Grant--in--Aid for Scientific Research (Grant No.~15K05217) 
from the Japan Society for the Promotion of Science (JSPS),
and a grant from Keio University.
\newcounter{firstbib}


\begin{thebibliography}{99}
\bibitem{oct1}N. Tanno, T. Ichikawa, A. Saeki,
 ``Lightwave Reflection Measurement'',
 Japanese Patent \#2010042 (1990).
\bibitem{oct2} D. Huang,  et al, 
``Optical coherence tomography''
% d (huang, d); swanson, ea (swanson, ea); lin, cp (lin, cp); schuman, js (schuman, js); stinson, wg (stinson, wg); chang, w (chang, w); hee, mr (hee, mr); flotte, t (flotte, t); gregory, k (gregory, k); puliafito, ca (puliafito, ca) 
Science 254, 1178-1181 (1991).
\bibitem{octReview1}
  A.F. Fercher,  W. Drexler, C.K. Hitzenberger, T. Lasser, 
  ``Optical coherence tomography - principles and applications'',
  Rep. Prog.  Phys. 66, 239---303 (2003).
\bibitem{sdoct1}A.F. Fercher, C.K.  Hitzenberger, G.  Kamp, S.Y. Elzaiat, 
  ``Measurement of intraocular distances by backscattering spectral interferometry'',
  Opt. Comm.	117, 43---48 (1995).
\bibitem{sdoct2} G. Hausler,  M.W.  Lindner, 
  ``Coherence radar'' and "spectral radar'' -- new tools for dermatological diagnosis'', J.
  Biomed. Opt. 3, 21-31 (1998).
\bibitem{Mitsui99} T. Mitsui,
``Dynamic range of optical reflectometry   with spectral interferometry'', 
  Jpn. J. Appl. Phys. 38, 6133---6137  (1999).
\bibitem{Leitgeb2003}R. Leitgeb, C.K. Hitzenberger, A. Fercher, 
  ``Performance of fourier domain vs. time domain optical coherence'
  tomography'', Opt. Exp. 11, 889---894 (2003).
\bibitem{Boer2003}J.F. de Boer et al, 
   % Cense, B
   % Park, BH
   % Pierce, MC
   % Tearney, GJ
   % Bouma, BE
  ``Improved signal-to-noise ratio in spectral-domain compared with
  time-domain optical coherence tomography'', 
  Opt. Lett.  28,  2067---2069 (2003).
\bibitem{octDoppler}
  X.J. Wang, T.E. Milner, J.S. Nelson, 
  ``Characterization of fluid-flow velocity by optical doppler tomography'',
  Opt. Lett. 20,  1337---1339 (1995).
\bibitem{octDoppler2}S. Makita et al, ``Optical coherence
  angiography'', Opt. Express, 14, 7821---7840 (2006).
\bibitem{octDopplerReview}A.Q. Zhang,  Q.Q. Zhang,  C.L. Chen, R.K.K Wang,
``Methods and algorithms for optical coherence
   tomography-based angiography: a review and comparison'',
  J. Biomed. Opt.  20,  100901 (2015).
  \bibitem {ripplonExp} D. Langevin (ed), ``Light scattering by liquid
    surfaces and complementary techniques'', Marcel Dekker, New York
    (1992).
\bibitem{Cicuta2004} P. Cicuta, L. Hopkinson, 
  ``Recent developments of surface light scattering as a tool for optical-rheology of polymer
   monolayers'', 
  Colloids and Surfaces A: Physicochem. Eng. Aspects 233,
  97---107 (2004).
  \bibitem{Sagis}%
    L.M.C. Sagis, 
    ``Dynamic properties of interfaces in soft matter:
    Experiments and theory'', 
    Rev. Mod. Phys. 83, 1367---1403 (2011).
\bibitem{Swanson92}E.A. Swanson, et al,
``High-speed optical coherence domain reflectometry'',
  Opt. Lett. 17, 151---153 (1992).
\bibitem{MA1}T. Mitsui, K. Aoki, 
`` Direct optical observations of surface
     thermal motions at sub-shot noise levels'', 
  Phys. Rev. E80,
  020602(R) (2009).
\bibitem{Shinetsu} Shin-Etsu Chemical Co., Ltd,
  \url{https://www.shinetsusilicone-global.com/catalog/index.shtml}.
\bibitem{CRC}W.M.  Haynes,  ``CRC Handbook of Chemistry and Physics, 92nd Edition'', 
CRC Press (Baton Rouge, 2011). 
  \bibitem {Levich} V.G. Levich, ``Physicochemical Hydrodynamics'',
    Prentice-Hall, Englewood Cliffs (1962).
  \bibitem {Bouchiat} M.-A. Bouchiat,  J. Meunier, 
    ``Power spectrum of
    fluctuations thermanny excited on free surface of a simple liquid'',
    J. de Phys. 32, 561---571 (1971).
  \bibitem{Jackle}J. Jackle, 
    ``The spectrum of surface waves on
     viscoelastic liquids of arbitrary depth'', 
    J. Phys. Cond. Matt. 10,
    7121---7131 (1998).
  \bibitem{MA2}T. Mitsui and K. Aoki, 
``Measurements of liquid surface
     fluctuations at sub-shot-noise levels with Michelson     interferometry'', 
    Phys. Rev. { E 87}, 042403 (2013).
\end{thebibliography}
\end{document}